\documentclass[namedreferences]{kluwer}
\usepackage{psfig}

\def \ap{Appl. Phys.}
\def \apj{Astrophys. J.}
\def \mnras{Mon. Not. Roy. Soc.}
\def \pra{Phys. Rev. A}
\def \prl{Phys. Rev. Lett.}
\def \jpcrd{J. Phys. Chem. Ref. Data}
\def \zpa{Z. Phys. A}

\def \da{\Delta\alpha/\alpha}
\def \citet{\inlinecite}
\def \citep{\cite}

\begin{document}

\begin{opening}         

\title{Does the fine structure constant vary? A detailed invest-igation
  into systematic effects}
\author{M.~T. \surname{Murphy}, J.~K. \surname{Webb},
V.~V. \surname{Flambaum}, S.~J. \surname{Curran}}
\runningauthor{Murphy et al.}
\runningtitle{Does $\alpha$ vary? Systematic effects}
\institute{School of Physics, University of New South Wales, Sydney
N.S.W. 2052\\Email: mim@phys.unsw.edu.au (MTM)}
\date{September 30, 2002}

\begin{abstract}
We have previously presented 5.7\,$\sigma$ evidence for a smaller $\alpha$
at redshifts $0.2 < z_{\rm abs} < 3.7$ from a sample of 128 Keck/HIRES
quasar absorption systems: $\Delta\alpha/\alpha = (-0.57 \pm 0.10) \times
10^{-5}$. A non-zero $\da$ manifests itself as a distinct pattern of shifts
in the measured absorption line wavelengths. The statistical error is now
small: we do detect small line shifts in the HIRES data. Whether these
shifts are due to systematic errors or due to real variation in $\alpha$ is
now the central question. Here we summarize the two potentially most
important systematic effects: atmospheric dispersion and isotopic abundance
evolution. Previously, these have been difficult to quantify/model but here
we find that neither of them can explain our results. Furthermore, the
HIRES spectra themselves contain no evidence for these effects. Independent
measurements of $\da$ with a different telescope and spectrograph are now
crucial if we are to rule out or confirm the present evidence for a
variable $\alpha$.
\end{abstract}
\keywords{ line: profiles -- instrumentation: spectrographs -- methods:
  data analysis -- techniques: spectroscopic -- quasars: absorption lines}

\end{opening}           

\section{Introduction}

In \citet{MurphyM_02a} and \citet{WebbJ_03b} we presented new evidence for
a smaller fine structure constant, $\alpha \equiv e^2/\hbar c$, in high
redshift quasar absorption clouds: $\da = (-0.57 \pm 0.10)\times
10^{-5}$. This preliminary result is the weighted mean from a
many-multiplet (MM) analysis of 128 absorption systems and represents a
significant improvement on our previously published result from 49 systems
\citep{MurphyM_01a,WebbJ_01a}: $\da = (-0.72 \pm 0.18)\times 10^{-5}$. Such
a potentially fundamental result requires extreme scrutiny, including a
thorough consideration of possible systematic errors. In
\citet{MurphyM_01b} we explored a wide range of possibilities: laboratory
wavelength errors, wavelength calibration errors, atmospheric dispersion
effects, unidentified interloping transitions, isotopic ratio and/or
hyperfine structure effects, intrinsic instrumental profile variations,
spectrograph temperature variations, heliocentric velocity corrections,
kinematic effects and large scale magnetic fields. All but two of these --
atmospheric dispersion effects and isotopic abundance variation -- were
shown to be negligible.

Here we present an updated analysis of these two effects. We also present
new results from a line removal test which constrains systematic errors
associated with individual transitions or species.

\section{Atmospheric dispersion effects}

\begin{figure}[t]
\centerline{\hbox{
    \vbox{
      \psfig{file=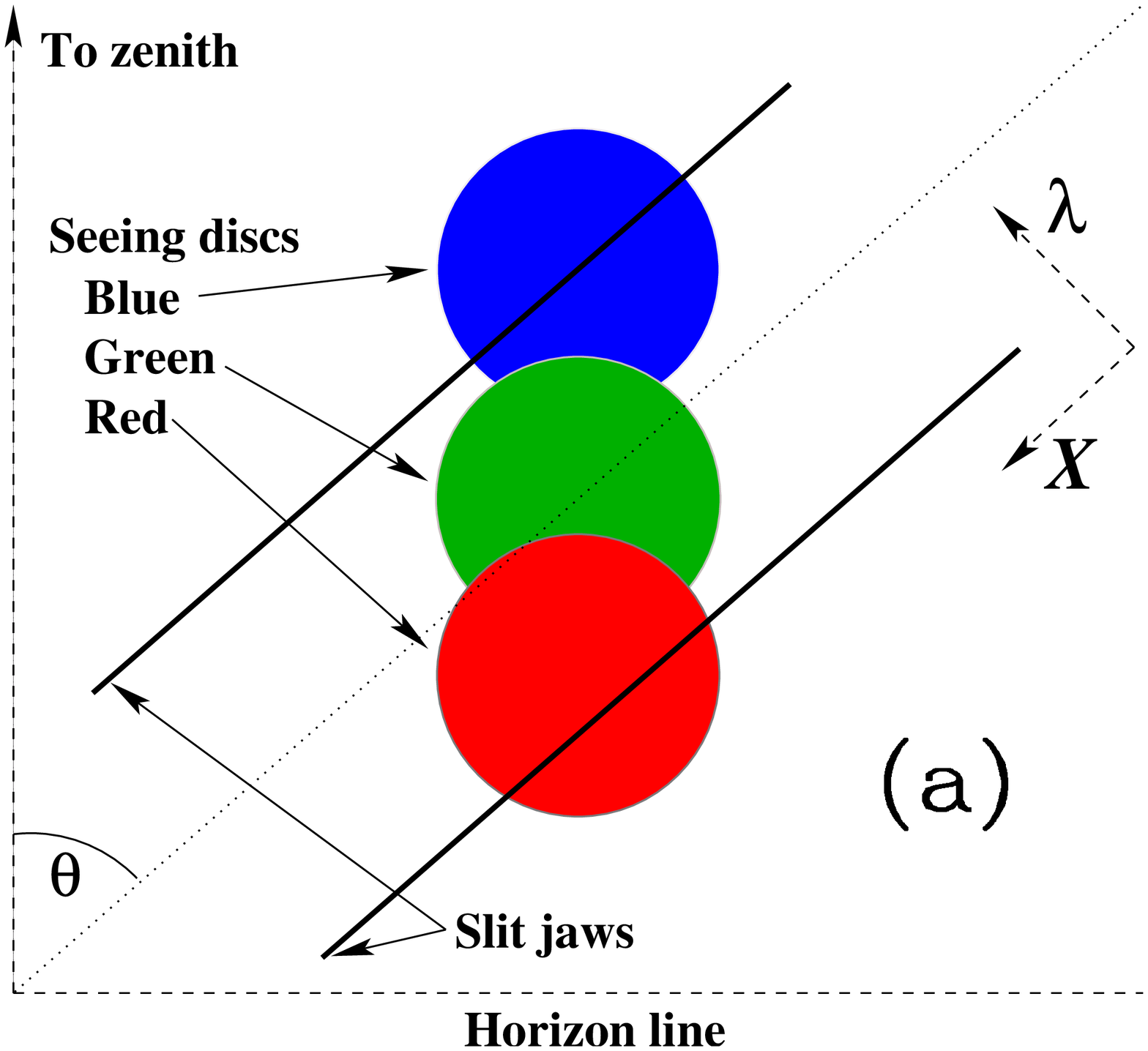,width=4.3cm}
      \psfig{file=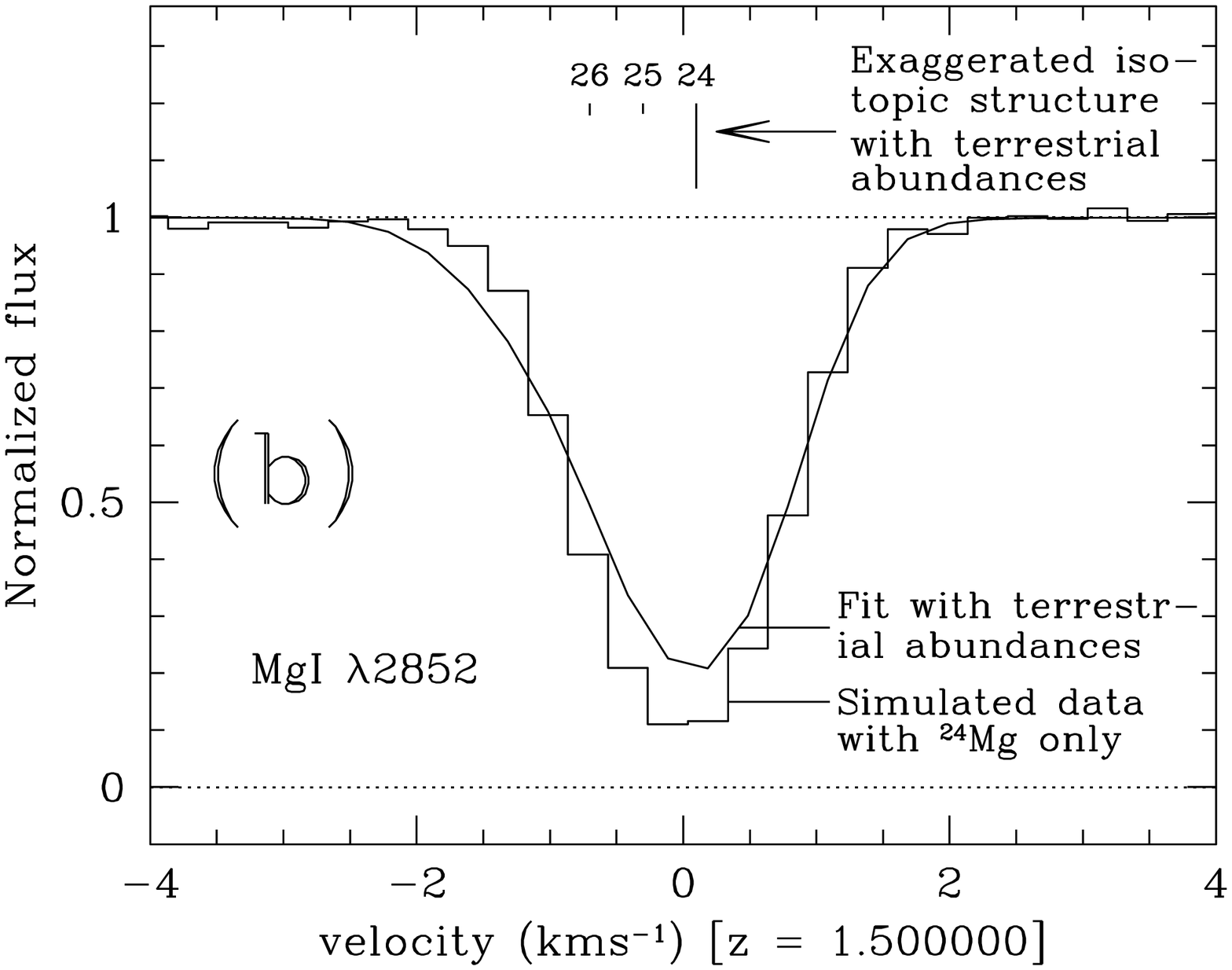,width=4.3cm}
    }
    \psfig{file=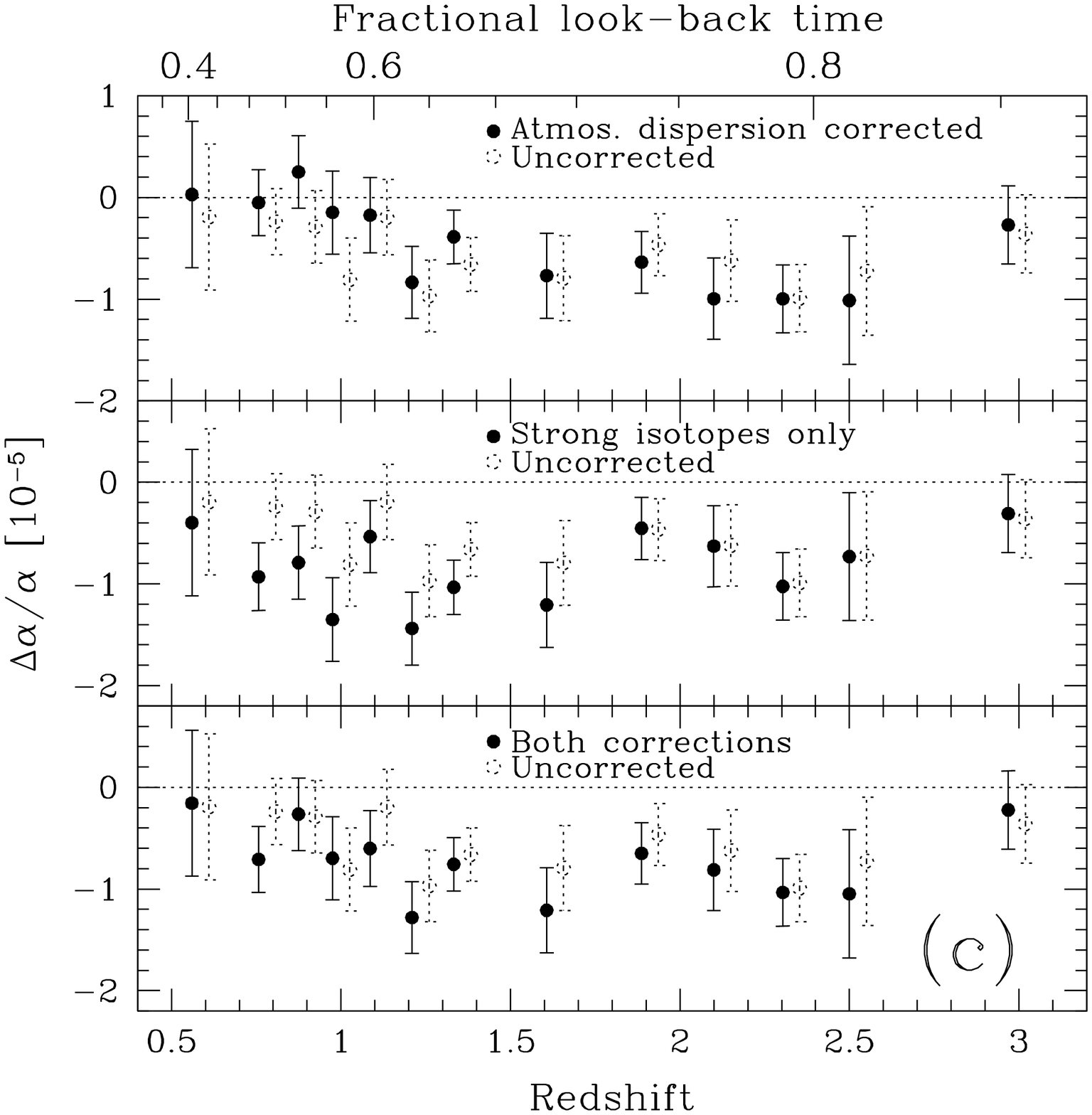,width=7.2cm}
  }
}
\caption{Summary of known, potentially significant systematic errors. {\bf
  (a)} Earth's atmosphere disperses seeing discs of different wavelength
  along the wavelength direction ($\lambda$) of the spectrograph slit,
  potentially compressing the quasar spectrum. The seeing discs are also
  truncated by the slit edges, leading to wavelength dependent IP
  asymmetry. {\bf (b)} Simulated Mg{\sc \,i} $\lambda$2852 absorption line
  (histogram) with only the $^{24}$Mg isotope. A fit to the data (solid
  line) with the full isotopic structure (tick-marks; separations
  exaggerated by a factor of 25), using the terrestrial isotopic ratios,
  induces an artificial line shift. {\bf (c)} The upper panel corrects
  $\da$ for atmospheric dispersion effects and the middle panel corrects
  for strong isotopic ratio evolution of $^{25,26}$Mg and
  $^{29,30}$Si. Both corrections are included in the lower panel.}
\label{fig:syserr}
\end{figure}

Fig. \ref{fig:syserr}a illustrates how atmospheric dispersion results in an
angular separation between different wavelengths as they enter the
spectrograph slit. This separation increases with increasing zenith angle,
$\xi$. For the particular optical design of the Keck/HIRES, the component
of angular separation across the slit (i.e. in the wavelength direction as
projected from the CCD) will cause a {\it compression} of the quasar
spectrum.  Such a compression will mimic $\da < 0$ for the Mg/Fe{\sc \,ii}
systems observed at $z_{\rm abs} < 1.8$, the effect being complicated at
high-$z$ by the diversity of $q$ coefficients\footnote{The $q$ coefficient
measures the sensitivity of a transition wavelength to changes in
$\alpha$. See \cite{DzubaV_02a} for details and updated values.} (see
Fig. \ref{fig:rmline}a). A wavelength dependent asymmetry will also be
introduced into the instrumental profile (IP) due to truncation of the
seeing discs by the slit edges. These effects will not be applied to the
thorium-argon (ThAr) calibration spectra since the ThAr lamp illuminates
the slit uniformly in all cases.

An image rotator is usually employed to hold the slit axis perpendicular to
the horizon as the quasar is tracked across the sky. This eliminates the
compression and IP asymmetry effects. Only 51 of our 128 absorption systems
were observed in this way; 77 systems were observed before August 1996 when
Keck/HIRES was fitted with an image rotator. However, the values of $\da$
for the 77 affected and 51 unaffected systems are consistent, with weighted
mean values $\da = (-0.54 \pm 0.14)\times 10^{-5}$ and $(-0.62 \pm
0.15)\times 10^{-5}$ respectively. We also expect to observe a correlation
between $\da$ and $\xi$, especially for the low-$z$ Mg/Fe{\sc \,ii} systems
(see Fig. \ref{fig:rmline}a). This expectation is confirmed in the
simulations below. However, no such correlation is observed in our quasar
data. Therefore, the data themselves suggest a negligible effect on $\da$
due to atmospheric dispersion effects.

Nevertheless, we used the atmospheric and telescope parameters for each
quasar observation to model the potential effect of atmospheric dispersion
on $\da$. The angular separation of two wavelengths entering the slit is
determined by the refractive index of air at the telescope (which depends
on temperature, pressure and relative humidity). The slit angle with
respect to the vertical, $\theta$, is equal to $\xi$ for the pre-rotator
Keck/HIRES. We applied the resulting wavelength shifts to a high S/N
simulation of our data and determined $\da$, giving a correction which we
apply to the values measured from the quasar spectra. We compare the
corrected values of $\da$ with the uncorrected values in
Fig. \ref{fig:syserr}c. Overall, our model of atmospheric dispersion has an
opposite effect on the low-$z$ systems compared to the high-$z$ systems: it
can not explain the observed deviation from $\da = 0$. Once we apply the
correction, the apparent evolution of $\alpha$ with look-back time is
enhanced.

\section{Isotopic ratio evolution}

Mg, Si, Cr, Fe, Ni and Zn have several stable isotopes and so the
transitions used in the MM method will have unresolved isotopic
structure. However, only the isotopic structures of the Mg lines have been
determined experimentally \citep{HallstadiusL_79a,DrullingerR_80a}. The Mg
and Si transitions involve similar energy levels and so, in
\citet{MurphyM_01b}, we estimated their isotopic structure by scaling the
Mg{\sc \,ii} $\lambda$2796 structure by the mass isotopic shift. We fitted
the Mg and Si absorption lines with these isotopic structures using the
terrestrial values of the isotopic abundances \citep{RosmanK_98a}. However,
Galactic observations of Mg \citep{GayP_00a} and theoretical models of Si
\citep{TimmesF_96a} in stars strongly suggest that only the $^{24}$Mg and
$^{28}$Si isotopes will exist in quasar absorption clouds with significant
abundances. Fig. \ref{fig:syserr}b illustrates the effect this would have
on a measured Mg{\sc \,i} $\lambda$2852 absorption line position (the
isotopic separations are exaggerated by a factor of 25).

To obtain an upper limit on the effect of isotopic ratio evolution on
$\da$, we refit the Mg and Si transitions using only the $^{24}$Mg and
$^{28}$Si isotopes, i.e. zero abundances of $^{25,26}$Mg and $^{29,30}$Si
in the absorption clouds. We compare these new, corrected values of $\da$
with the uncorrected values in Fig. \ref{fig:syserr}c (middle panel). As
expected, the potential effect is most prominent in the low-$z$ Mg/Fe{\sc
\,ii} systems. Note that $\da$ is more negative after the correction:
isotopic ratio evolution can not explain our results. Although the isotopic
structures of the other transitions in our analysis are not known, they
should be more compact since the mass isotopic shift is inversely
proportional to the square of atomic mass. The line removal test described
below largely rules out such strong effects.

\section{Line removal tests}

\begin{figure}[t]
\centerline{\hbox{
    \vbox{
      \psfig{file=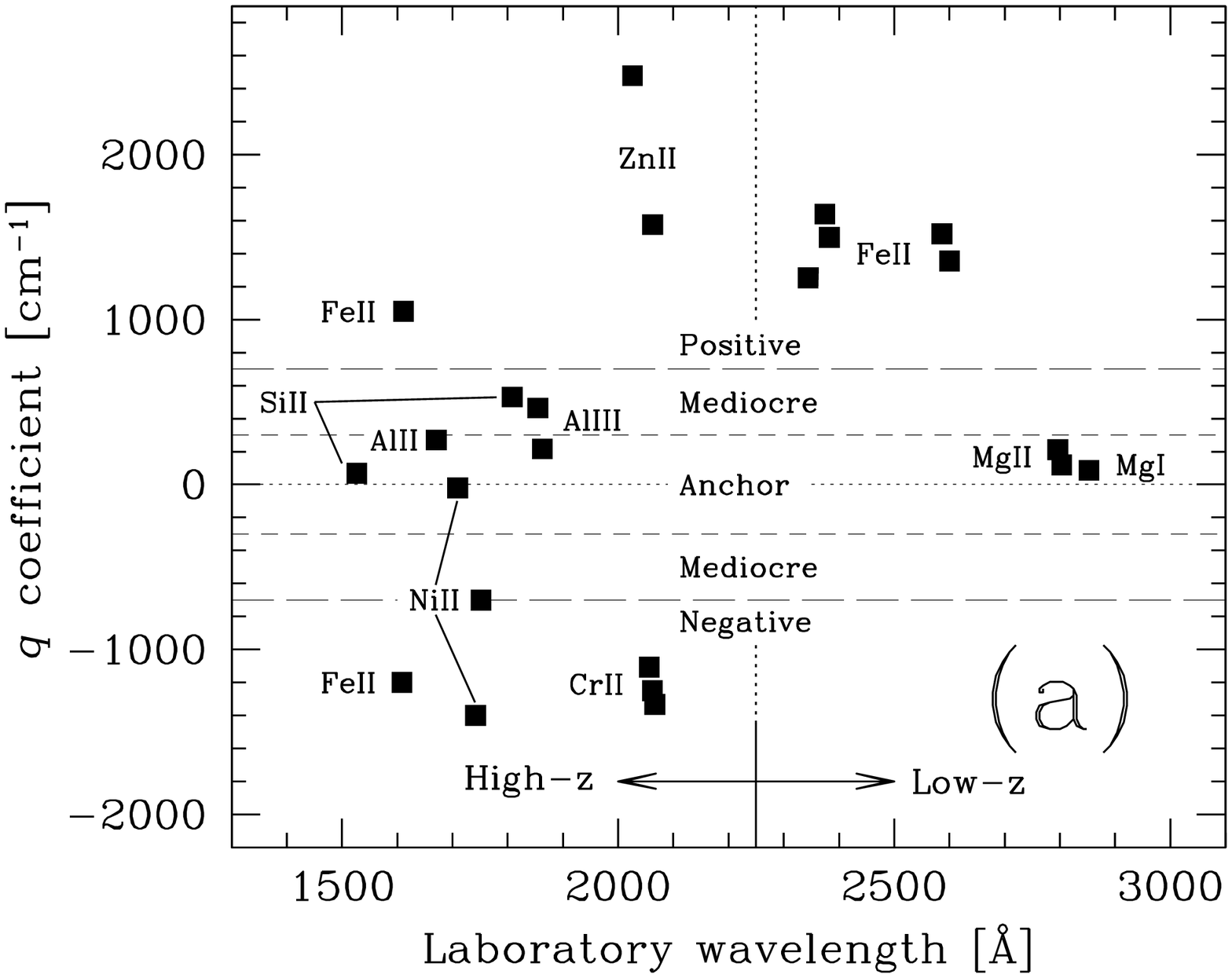,width=5.3cm}
      \psfig{file=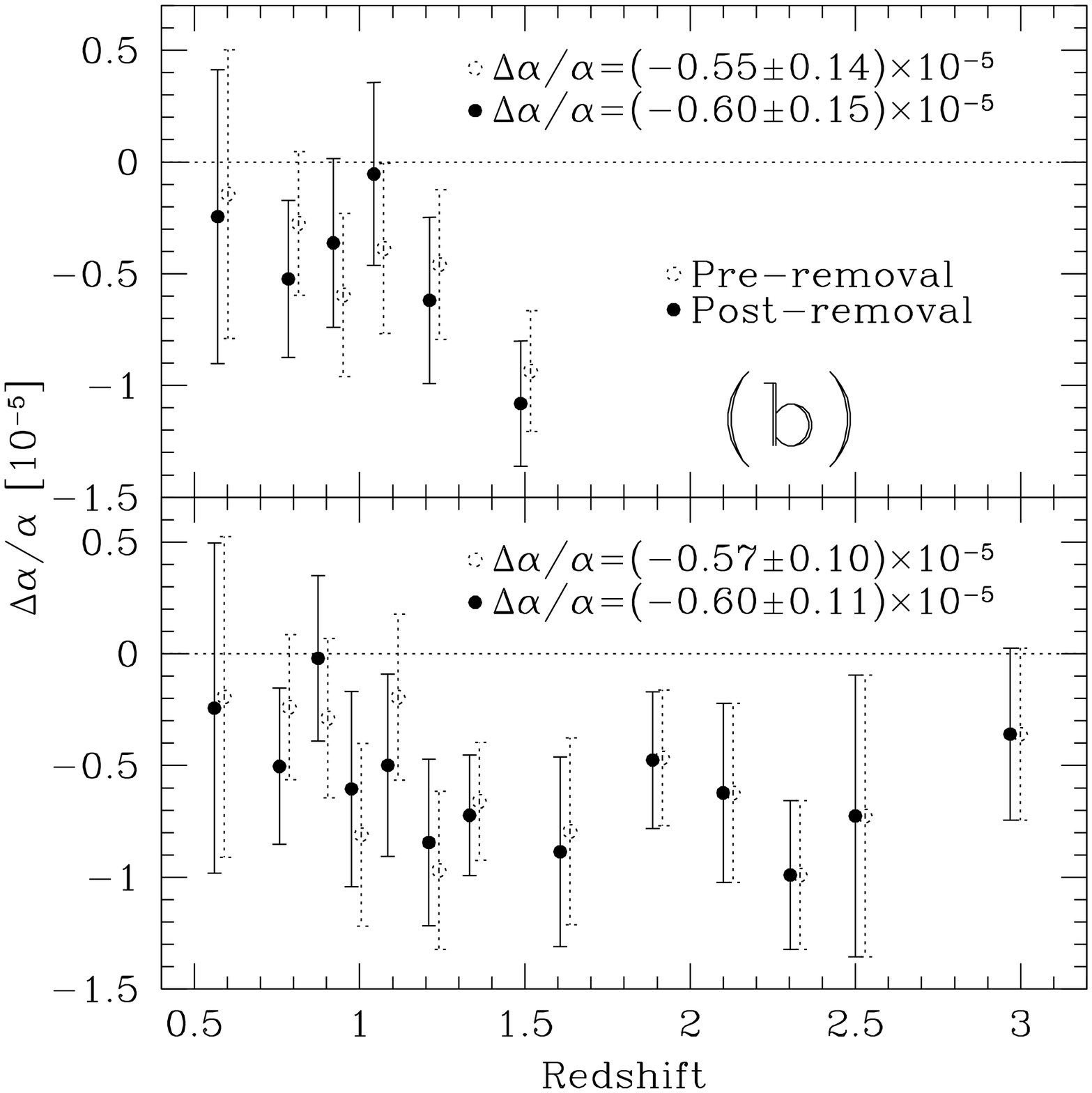,width=5.4cm,height=5.0cm}
    }
    \psfig{file=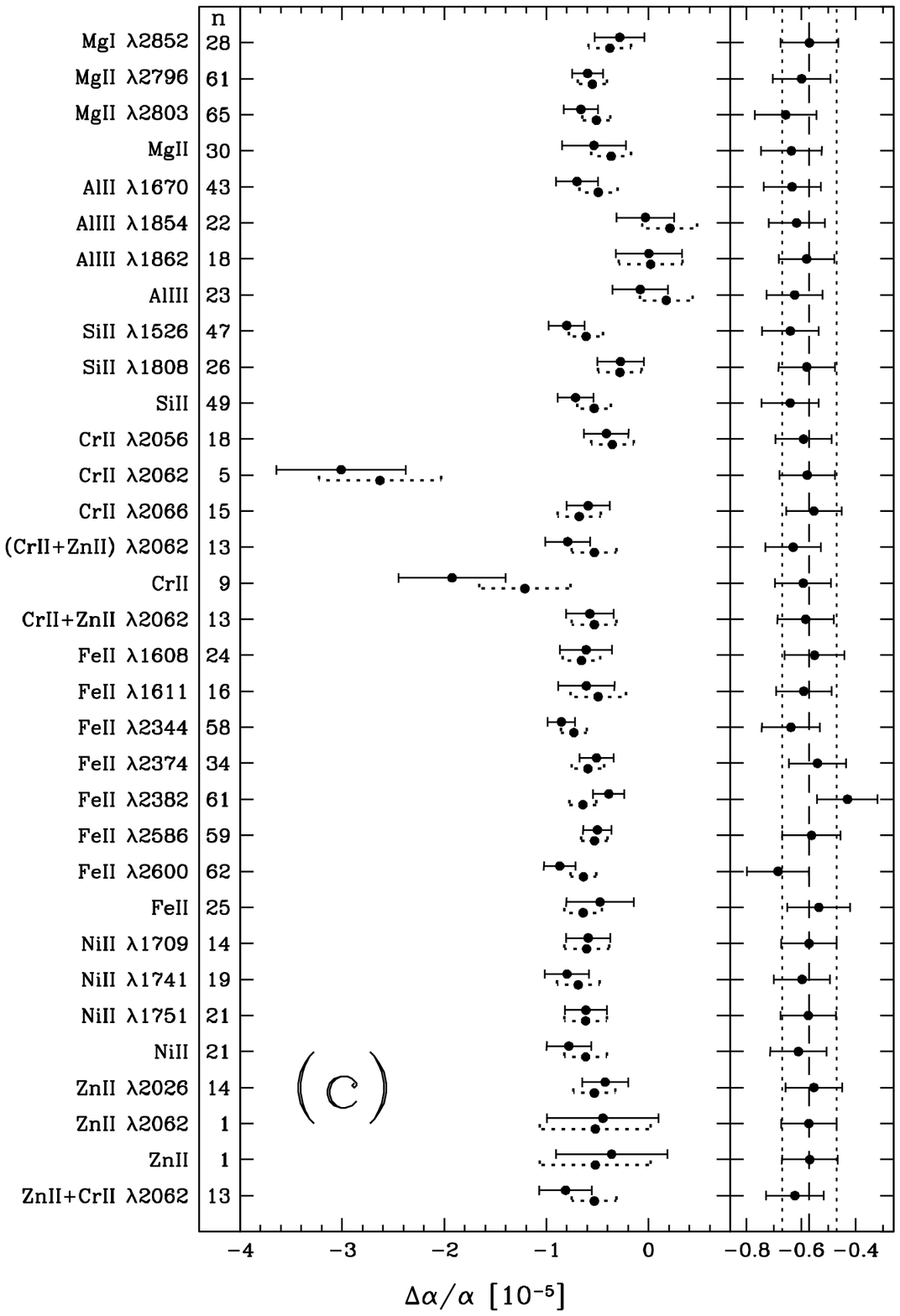,width=6.3cm}
  }
}
\caption{{\bf (a)} Distribution of $q$ coefficients for all transitions in
    the MM method. The simple arrangement for the low-$z$ Mg/Fe{\sc \,ii}
    systems indicates susceptibility to low-order distortions of the
    wavelength scale. {\bf (b)} Removing the Mg{\sc \,i} $\lambda$2796 line
    is possible for $n=61$ systems. The effect of removal on these systems
    (upper panel) and on the entire sample (lower panel) is small,
    indicating negligible systematic errors associated with Mg{\sc \,i}
    $\lambda$2796. {\bf (c)} Full line removal results. The
    transition/species removed is given on the vertical axis along with the
    number of systems, $n$, where removal was possible. The left panel
    compares the weighted mean value of $\da$ before (dotted error bar) and
    after (solid error bar) removal for these $n$ systems. The right panel
    shows the effect of removal on $\da$ for the entire sample,
    i.e. including systems where the transition/species could not be
    removed.}
\label{fig:rmline}
\end{figure}

A non-zero $\da$ manifests itself as a distinct pattern of line shifts
(Fig. \ref{fig:rmline}a). Thus, with a large set of fitted transitions, one
or more transitions can be removed from a given system and $\da$ will
remain well-constrained. In the absence of systematic errors associated
with the transition removed or with the wavelength scale of the quasar
spectra, $\da$ should not change systematically for all systems. For
example, removing a single transition (e.g. Mg{\sc \,ii} $\lambda$2796)
from our fits to the quasar data directly probes possible systematic errors
due to line blending, isotopic/hyperfine structure effects and errors in
the laboratory wavelengths. Removing entire species (e.g. all transitions
of Si{\sc \,ii}) also tests for isotopic/hyperfine structure effects.

Given a set of fitted transitions, we only remove transitions that {\it
can} be removed. To clarify this, consider removing Mg{\sc \,ii}
$\lambda$2796 (Fig. \ref{fig:rmline}b). For the low-$z$ Mg/Fe{\sc \,ii}
systems, a well-constrained value of $\da$ can only be obtained when at
least one Mg line is fitted since the difference between the $q$
coefficients for the various Fe{\sc \,ii} transitions is small (see
Fig. \ref{fig:rmline}a). If the only available anchor line is Mg{\sc \,ii}
$\lambda$2796 then we can not remove it from the system. Similarly, we only
remove the entire Mg{\sc \,ii} species if the Mg{\sc \,i} $\lambda$2852
line is fitted.

In Fig. \ref{fig:rmline}c we present line removal results for all
transitions and species. The left panel shows the overall effect line
removal has on the $n$ systems where line removal is possible (cf. upper
panel of Fig. \ref{fig:rmline}b). The right panel shows the effect on the
entire sample (cf. lower panel of Fig. \ref{fig:rmline}b). We see no
evidence for systematic errors associated with particular transitions or
species.

\section{Conclusions}

We have extended our previous analysis of systematic errors
\citep{MurphyM_01b} to include the new, significantly augmented sample of
128 absorption systems outlined in \citet{MurphyM_02a} and
\citet{WebbJ_03b}. We cannot explain the observed value of $\da = (-0.57
\pm 0.10)\times 10^{-5}$ with known systematic errors. Here we have
outlined our analysis of the two most important of these: atmospheric
dispersion effects and possible isotopic ratio evolution. The former is an
instrumental effect that should have affected 77 of the 128 absorption
systems. However, the data themselves suggest that atmospheric dispersion
had a negligible effect on $\da$. Furthermore, modelling indicates that the
resulting low-order distortion of the wavelength scale should have shifted
the low- and high-$z$ values of $\da$ in opposite senses, clearly at odds
with our observations. An upper limit on the effect on $\da$ of strong
isotopic ratio evolution was obtained by assuming zero isotopic abundances
for $^{25,26}$Mg and $^{29,30}$Si in the quasar absorption clouds. Although
the isotopic structures of the other transitions used in the MM method are
not known, strong isotopic ratio evolution for these elements is strongly
constrained by the line removal test (Fig. \ref{fig:rmline}c).

All 128 absorption systems were observed with the Keck/HIRES. Although we
have ruled out many instrumental systematic errors, independent
measurements of $\da$ must be carried out with a different telescope and
spectrograph. Using the MM method to analyse VLT and Subaru spectra will be
a crucial check on possible systematic errors.

%\bibliography{references}
%\bibliographystyle{klunamed}

\end{document}